\documentclass[aps,showpacs,superscriptaddress,floatfix,onecolumn]{revtex4}
\usepackage{epsfig}\usepackage{amsmath}
\usepackage{amsmath,amsfonts,amssymb,graphics,graphicx,epsfig,color,times,bbm}
\newcommand{\id}{{\sf 1 \hspace{-0.3ex} \rule{0.1ex}{1.52ex}\rule[-.01ex]{0.3ex}{0.1ex}}}
\usepackage{mathptmx}

\DeclareFontFamily{OML}{caligr}{}\DeclareFontShape{OML}{caligr}{m}{it}{%
  <-5>sub * cmr/m/n <5-7>rsfs5 <7-10>rsfs7 <10->rsfs10}{}

\begin{document}

%\preprint{APS/123-QED}

\title{Single mode two-channel cavity QED}% Force line breaks with \\
\author{F. L. Semi\~ao}
\affiliation{Departamento de F\'isica, Universidade Estadual de Ponta Grossa - Campus Uvaranas, 84030-900 Ponta Grossa, Paran\'a, Brazil}

\begin{abstract}
In this short communication, a new type of two-channel cavity QED model is derived. Two-channel models are important for they often lead to quantum interference phenomena. The previous models relied on the use of two or more modes of the quantized electromagnetic field, partially because of energy and parity restrictions. As it is shown in this work, such restrictions may be overcome with the use of properly chosen configurations of atomic levels and the aplication of classical external fields. Competing one- and two-photon processes involving one single mode may be obtained.  
\end{abstract}

\maketitle

The interaction between electromagnetic fields and matter plays a central role in physics. This topic has been studied from many different points of view and approaches. From high energy physics to cryogenic physics, and from single atoms to bulk matter, the knowledge of light-matter interaction mechanisms is of fundamental importance. The quantum description of these interactions in the low energy domain (non-relativistic) is the object of quantum optics. In this context, many successful theoretical models have been proposed and experimentally implemented. One of such models is the well-known Jaynes-Cummings model \cite{JCM} which, among many interesting features, predicts the occurrence of collapses and revivals of coherence in the dynamics of a two-level atom in interaction with a single mode of the quantized light field \cite{revivals}. The coupling between an atom and a mode of the quantized electromagnetic field is greatly enhanced in a cavity. For cavities with a high quality factor Q (microwave regime) or finesse \textsl{F} (optical regime), the atom emits preferably into one of the cavity modes. This special arrangement allows for experimental investigation of fundamental photon-atom interactions, and it is called cavity quantum electrodynamics (cavity QED). An experimental review of this topic may be found in \cite{review}.

Among the several cavity QED models, there is a special class which deals with competing processes involving atomic  transitions. These multi-channel cavity QED models include pump-probe processes such as Stokes and anti-Stokes emission with classical \cite{chclass} or quantum pump \cite{chquan}, and competing nonlinear atom-multimode couplings \cite{c1,onetwo,c2,c3}. The dynamics in those models often reveals signatures of quantum interference. The situation is analogous to the double-slit experiment with particles. In a two-channel model, the atom may pass from one state to another through two different \emph{paths}. Interference between these two different amplitudes of probability generally leads to interesting phenomena such as quantum beats \cite{qb,scully} and coherent trapping \cite{scully,ct}. These multi-channel models have also motivated the idea of trapping field states \cite{c2,tfs}. Field states can trap the atomic population in the ground state via quantum interference between two or more channels. A common feature of all these multi-channel models is the use of two or more modes of the quantized field. A natural question would be, is it possible to construct multi-channel cavity QED models employing just one cavity mode? This paper is intended to provide a positive answer to this question. 

\begin{figure}[ht]
\centering
\includegraphics[scale=0.9]{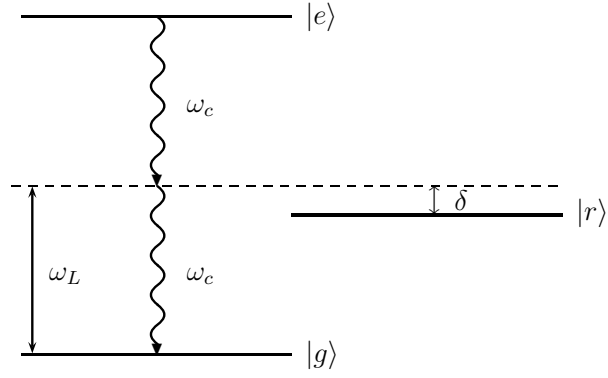}
\caption{Schematic diagram of the three-level atom in interaction with a quantized field (angular frequency $\omega_c$) and a classical laser field (angular frequency $\omega_L$). For sufficiently large $\delta$, the level $|r\rangle$ may be eliminated, and two-photon transitions between $|e\rangle$ and $|g\rangle$ are enhanced. Due to the presence of the laser, there is no energy limitations for the atom also make one-photon transitions between such levels.}
\label{fig1}
\end{figure}

In this short communication, a single mode two-channel cavity QED model involving atomic one- and two-photon transitions is derived. At first, the idea of having such a situation would seem to be forbidden for energy conservation and parity reasons. However, the assistance of an external laser field and the proper choice of the atomic levels can lead to the desired interaction. The physical system considered here is composed of a three-level atom which simultaneously interacts with a single-mode quantized cavity field and an external laser field, as depicted in Fig.(\ref{fig1}). The system Hamiltonian is given by
\begin{eqnarray}
\hat{H}&=&\hbar\omega_c\hat{a}^\dag\hat{a}+ E_e\sigma_{ee}+ E_r\sigma_{rr}+ E_g\sigma_{gg}+\hbar\Omega(\sigma_{gr}e^{i\omega_Lt}+\sigma_{rg}e^{-i\omega_Lt})+\hbar(g_1\sigma_{gr}+g_2\sigma_{re})\hat{a}^\dag+\hbar(g_1\sigma_{rg}+g_2\sigma_{er})\hat{a},\label{original}
\end{eqnarray}   
where $\omega_c$ and $\omega_L$ are the frequencies of the cavity and laser fields, respectively, $E_i$ is the energy of electronic level $i$, $\Omega$ is the Rabi frequency for the atom-laser coupling, and $g_1$ ($g_2$) is the coupling constant for the quantized field induced transitions $|g\rangle\leftrightarrow|r\rangle$ ($|r\rangle\leftrightarrow|e\rangle$). The coupling constants are assumed to be real for the sake of simplicity. 
%%%%%%%%%%%%%%%%%%%%%%%%%%%%%%%%%%%%%%%%%%%
%%%%%%%%%%%%%%%%%%%%%%%%%%%%%%%%%%%%%%%%%%% 

In the interaction picture, Hamiltonian (\ref{original}) reads
\begin{eqnarray}
\label{Hip}
\hat{H}_I(t)&=&\hbar\Omega(\sigma_{gr}e^{i\delta t}+\sigma_{rg}e^{-i\delta t})+\hbar g_1(\sigma_{gr}\hat{a}^\dag e^{i\delta t}+\sigma_{rg}\hat{a}e^{-i\delta t})+\hbar g_2(\sigma_{re}\hat{a}^\dag e^{-i\delta t}+\sigma_{er}\hat{a}e^{i\delta t}),
\end{eqnarray}
where $\delta$ is the detuning involving the level $|r\rangle$. The above interaction picture Hamiltonian is a special instance of 
\begin{eqnarray}
\hat{H}_I(t)=\hbar\sum_{k=1}^N\lambda_k(\hat{A}_ke^{i\delta_k t}+\hat{A}_k^\dag e^{-i\delta_k t}),\label{gH}
\end{eqnarray}
which describes many important systems in quantum optics. The case $\delta_k=\delta$, for all $k$, is considered in the following derivation, but the reader should not have trouble to develop the same line of reasoning for the general case. The method presented here to obtain effective Hamiltonians in dispersive regimes is a generalization of the methods described in \cite{method}.

Consider the time (ordered) evolution operator for a time-dependent Hamiltonian
\begin{eqnarray}
\hat{U}_I(t)=\id-\frac{i}{\hbar}\int_0^t dt^{'}\hat{H}_I(t^{'})-\frac{1}{\hbar^2}\int_0^t dt^{'}\hat{H}_I(t^{'})\int_0^{t^{'}} dt^{''}\hat{H}_I(t^{''})+...\,.\label{umethod}
\end{eqnarray}
Simple analysis of the above expression reveals that it involves a crescent power series in $\lambda_k/\delta$ with time-dependent coefficients. In the dispersive regime $\delta\gg \lambda_k$, for all $k$, just terms up to second order should be considered. First and second order contributions will be carefully studied now for the Hamiltonian (\ref{gH}).

The first order contribution in (\ref{umethod}) using (\ref{gH}) reads
\begin{eqnarray}
\int_0^t dt^{'}\hat{H}_I(t^{'})=\hbar\sum_{k=1}^N\frac{\lambda_k}{i\delta}[(1-e^{-i\delta t})\hat{A}_k^\dag-(1-e^{i\delta t})\hat{A}_k],\label{fo}
\end{eqnarray}
and the second order contribution is given by
\begin{eqnarray}
\int_0^t dt^{'}\hat{H}_I(t^{'})\int_0^{t^{'}} dt^{''}\hat{H}_I(t^{''})&=&\hbar^2\int_0^t dt^{'}\sum_{k,j=1}^N\frac{ \lambda_j\lambda_k}{i\delta}[(e^{-i\delta t^{'}}-e^{-2i\delta t^{'}})\hat{A}_j^\dag\hat{A}_k^\dag-(e^{-i\delta t^{'}}-1)\hat{A}_j^\dag\hat{A}_k+(e^{i\delta t^{'}}-1)\hat{A}_j\hat{A}_k^\dag\nonumber\\ &&-(e^{i\delta t^{'}}-e^{2i\delta t^{'}})\hat{A}_j\hat{A}_k].
\end{eqnarray}
After integration, the oscillating terms will give rise to multiplicative factors proportional to $(\lambda_j\lambda_k/\delta)^2$, whereas the others will appear multiplied by $(\lambda_j\lambda_k/\delta)$. In the dispersive regime, we can safely drop the former to get
\begin{eqnarray}\label{so}
\int_0^t dt^{'}\hat{H}_I(t^{'})\int_0^{t^{'}} dt^{''}\hat{H}_I(t^{''})=\hbar^2\sum_{k,j=1}^N\frac{\lambda_j\lambda_k t}{i\delta}[\hat{A}_j^\dag\hat{A}_k-\hat{A}_j\hat{A}_k^\dag]=\hbar^2\sum_{j=1}^N\frac{\lambda_j^2t}{i\delta}[\hat{A}_j^\dag,\hat{A}_j]+\hbar^2\sum^N_{k,j=1\atop{}k\neq{}j}\frac{\lambda_j\lambda_kt}{i\delta}[\hat{A}_j^\dag,\hat{A}_k].
\end{eqnarray}
The Stark-shifts arise from the first term whereas the interaction between the subsystems comes from the double summation. The second order contribution (\ref{so}) is much more important to $\hat{U}_I(t)$ than the first order one (\ref{fo}), because the former involves terms linear in time whereas the latter involves terms that are just oscillatory or constant in time. Consequently, just the second order terms are retained, and the time evolution operator (\ref{umethod}) assumes the form
\begin{eqnarray}
\hat{U}_I(t)\approx\id-it\left[\sum_{j=1}^N\frac{\lambda_j^2}{\delta}[\hat{A}_j,\hat{A}_j^\dag]+\sum^N_{k,j=1\atop{}k\neq{}j}\frac{\lambda_j\lambda_k}{\delta}[\hat{A}_j,\hat{A}_k^\dag]\right].
\end{eqnarray} 
One can now easily identify the interaction picture time-independent effective Hamiltonian by comparing the above expression with $\hat{U}_I(t)\approx\id-i t\hat{H}_I^{\rm eff}/\hbar$. Clearly,
\begin{eqnarray}
\hat{H}_I^{\rm eff}=\hbar\sum_{j=1}^N\frac{\lambda_j^2}{\delta}[\hat{A}_j,\hat{A}_j^\dag]+\hbar\sum^N_{k,j=1\atop{}k\neq{}j}\frac{\lambda_j\lambda_k}{\delta}[\hat{A}_j,\hat{A}_k^\dag].
\end{eqnarray}
For the physical system considered in this paper, the above formula is to be used with $N=3$, $\lambda_1=g_1$, $\lambda_2=g_2$, $\lambda_3=\Omega$,  $\hat{A}_1=\sigma_{gr}\hat{a}^\dag$, $\hat{A}_2=\sigma_{er}\hat{a}$, and $\hat{A}_3=\sigma_{gr}$. For an atom that is not initially in $|r\rangle$, it results in the interaction picture Hamiltonian 
\begin{eqnarray}
\hat{H}_I^{\rm eff}=\hat{H}_{\rm{Stark}}+\hat{H}_{\rm{1ph}}+\hat{H}_{\rm{2ph}}+\hat{H}_{\rm{D}},\label{Ht}
\end{eqnarray}
where
\begin{eqnarray}
\hat{H}_{\rm{Stark}}=\hbar\frac{g_1^2}{\delta}\hat{a}^\dag\hat{a}\sigma_{gg}+\hbar\frac{g_2^2}{\delta}\hat{a}\hat{a}^\dag\sigma_{ee}+\hbar\frac{\Omega^2}{\delta}\sigma_{gg}
\end{eqnarray}
are the Stark-shifts induced by the fields,
\begin{eqnarray}
\hat{H}_{\rm{1ph}}=\hbar\frac{\Omega g_2}{\delta}(\sigma_-\hat{a}^\dag+\hat{a}\sigma_+)
\end{eqnarray}
is the ordinary one-photon Jaynes-Cummings Hamiltonian $[\sigma_+=(\sigma_-)^\dag\equiv\sigma_{eg}]$,
\begin{eqnarray}
\hat{H}_{\rm{2ph}}=\hbar\frac{g_1g_2}{\delta}(\sigma_-\hat{a}^\dag{}^2+\hat{a}^2\sigma_+)
\end{eqnarray}
is a Hamiltonian which accounts for atomic two-photon transitions, and
\begin{eqnarray}
\hat{H}_{\rm{D}}=\hbar\frac{\Omega g_1}{\delta}(\hat{a}^\dag+\hat{a})\sigma_{gg}\label{D}
\end{eqnarray}
is a displacement term controlled by the state of the atom.

The Hamiltonian (\ref{Ht}) presents a quite remarkable competing process. An atom initially prepared in $|e\rangle$, for instance, may emit either one or two photons into \textsl{the same} cavity mode. In the absence of the classical external field, and the properly choice of the atomic level configuration, such a situation would be clearly forbidden for energy conservation and parity reasons. Here, the single mode implementation is assisted by the classical field through the intermediate state $|r\rangle$, see Fig.({\ref{fig1}}). As a practical example, consider the typical cavity QED two-photon maser configuration using $^{85}$Rb Rydberg atoms \cite{HarocheExp,HarocheTh}. In this real physical setup, $|e\rangle$ and $|g\rangle$ correspond to the 40$S_{1/2}$ and 39$S_{1/2}$ states of $^{85}$Rb, respectively, and the intermediate state $|r\rangle$ corresponds to the state 39$P_{3/2}$. Therefore, the states $|e\rangle$ and $|g\rangle$ have the same parity making direct transitions between them quite unlikely in microwave cavities. However, the opposite-parity state $|r\rangle$ couples to $|e\rangle$ and $|g\rangle$ by dipole-allowed electric transitions with very large electric dipole moments. Indeed, according to the experiments reported in \cite{HarocheExp}, $g_1\approx g_2\equiv g$, with $g=7\times 10^5$ s$^{-1}$. What makes this particular experimental setup \cite{HarocheExp} suitable for the implementation of the ideas presented here is the appropriate magnitude of $\delta$. In this case, the small $|\delta|\approx 2.45\times 10^8$ s$^{-1}$ leads to $g^2/\delta\approx 6\times 10^3$ s$^{-1}$ which is more than enough to observe the second order effects described in this paper. Of course, direct application of laser beams is not possible in the microwave domain, but such large wavelengths may me achieved with two lasers in Raman configuration. Of course, the scheme shown in Fig.(\ref{fig1}) and the interaction Hamiltonian (\ref{Ht}) are valid in the optical regime as well. 

It should be remarked that simultaneous one- and two-photon transitions involving two different field modes has already been proposed and studied \cite{onetwo}. The present paper goes a step further bringing another completely different physical situation where such transitions take place with just one cavity mode.  The resulting Hamiltonian (\ref{Ht}) must contain new results which are related to quantum interference due to competing processes taking place with just one cavity mode. Besides, the role of the term (\ref{D}), which comes from the injection of energy from the laser into the system, may also lead to interesting consequences. However, the complicated form of (\ref{Ht}) renders analytical progress or numerical investigation a challenging task to be investigated elsewhere. 

In summary, this paper dealt with the subject of two-channel cavity QED, and a new competing process involving just one mode of the electromagnetic field was presented. Further investigation of the complicated dynamics of the system may reveal new phenomena related to quantum interference. It is expected that this short communication triggers future studies of the rich possibilities contained in this physically feasible new cavity QED model.

%%%%%%%%%%%%%%%%%%%%%%%%%%%%%%%%%%%%%%%%%%%%%%%%%%%%%%%%%%%%%%%%%%%%%%%%%%%%%%%%%%%%%%%%%%%%%%%%%%%%%%%%%%%%%%%%%%%%%%
\begin{acknowledgments}
I am thankful to K. Furuya and E. A. Chagas for inspiring discussions, and I also wish to thank P. P. Munhoz for the careful reading of this manuscript.
\end{acknowledgments}%%%%%%%%%%%%%%%%%%%%%%%%%%%%%%%%%%%%%%%%%%%%%%%%%%%%%%%%%%%%%%%%%%%%%%%%%%%%%%%%%%%%%%%%%%%%%%%%%%%%%%%%%%%%%%%%%%%%%%

%%%%%%%%%%%%%%%%%%%%%%%%%%%%%%%%%%%%%%%%%%%%%%%%%%%%%%%%%%%%%%%%%%%%%%%%%%%%%%%%%%%%%%%%%%%%%%%%%%%%%%%%%%%%%%%%%%%%%%%

%%%%%%%%%%%%%%%%%%%%%%%%%%%%%%%%%%%%%%%%%%%%%%%%%%%%%%%%%%%%%%%%%%%%%%%%%%%%%%%%%%%%%%%%%%%%%%%%%%%%%%%%%%%%%%%%%%%%%%%

\end{document}